\documentclass[prc,floatfix,groupedaddress,nofootinbib,showpacs,preprintnumbers,
amsmath,amssymb,amsfonts,superscriptaddress,widetable] {revtex4-1}
\usepackage{graphicx}% Include figure files
\usepackage{dcolumn}%Align table columns on decimal point
\usepackage{mathrsfs,amsmath,amssymb}
\usepackage{bm}% bold math
\usepackage{graphicx}% Include figure files
\usepackage{dcolumn}% Align table columns on decimal point
\usepackage{bm}% bold math
\usepackage{array}
\usepackage{bigstrut}
\usepackage[usenames]{color}

\begin{document}

\title{Validating neural-network refinements of nuclear mass models}
\author{R. Utama\footnote{Present address: Cold Spring Harbor Laboratory, NY 11724}}
\email{utama@cshl.edu} 
\affiliation{Department of Physics, Florida State University, Tallahassee, FL 32306} 
\author{J. Piekarewicz}
\email{jpiekarewicz@fsu.edu}
\affiliation{Department of Physics, Florida State University, Tallahassee, FL 32306}
\date{\today}
\begin{abstract}
\begin{description}
 \item[Background]  Nuclear astrophysics centers on the role of nuclear physics in 
  	                        the cosmos. In particular, nuclear masses at the limits of   
	                        stability are critical in the development of stellar structure and 
	                        the origin of the elements. 

\item[Purpose]        To test and validate the predictions of recently refined nuclear 
	                       mass models against the newly published AME2016 compilation.

\item[Methods]        The basic paradigm underlining the recently refined nuclear mass 
	                       models is based on existing state-of-the-art models that are  
	                       subsequently refined through the training of an artificial neural network. 
	                       Bayesian inference is used to determine the parameters of the 
	                       neural network so that statistical uncertainties are provided for
	                       all model predictions. 

\item[Results]	      We observe a significant improvement in the Bayesian 
	                       Neural Network (BNN) predictions relative to the corresponding 
	                       ``bare" models when compared to the nearly 50 new masses 
	                       reported in the AME2016 compilation. Further, AME2016 
	                       estimates for the handful of impactful isotopes in the determination 
	                       of $r$-process abundances are found to be in fairly good 
	                       agreement with our theoretical predictions. Indeed, the BNN-improved 
	                       Duflo-Zuker model predicts a root-mean-square deviation relative 
	                       to experiment of $\sigma_{\rm rms}\!\simeq\!400$\,keV.                        
	                      
\item[Conclusions]  Given the excellent performance of the BNN refinement in confronting
                                the recently published AME2016 compilation, we are confident of its
                                critical role in our quest for mass models of the highest quality. Moreover, 
                                as uncertainty quantification is at the core of the BNN approach, the 
                                improved mass models are in a unique position to identify those nuclei 
                                that will have the strongest impact in resolving some of the outstanding 
                                questions in nuclear astrophysics.                                      
\end{description}
\end{abstract} 
\maketitle

%%%%%%%%%%%%%%%%%%%%%%%%%%%%%%%%
\section{Introduction}
\label{intro}
%%%%%%%%%%%%%%%%%%%%%%%%%%%%%%%%

As articulated in the most recent US long-range plan 
for nuclear science\,\cite{LongRangePlan} ``nuclear 
astrophysics addresses the role of nuclear physics in our 
universe", particularly in the development of structure and 
on the origin of the chemical elements. In this context, 
fundamental nuclear properties such as masses, radii, and 
lifetimes play a critical role. However, knowledge of these 
nuclear properties is required at the extreme conditions 
of density, temperature, and isospin asymmetry found in 
most astrophysical environments. Indeed, exotic nuclei
near the drip lines are at the core of several fundamental 
questions driving nuclear structure and astrophysics 
today: \emph{what are the limits of nuclear binding?},
\emph{where do the chemical elements come from?},
and \emph{what is the nature of matter at extreme 
densities?}\,\cite{LongRangePlan,national2012Nuclear,
QuarksCosmos:2003}.

Although new experimental facilities have been commissioned 
with the aim of measuring nuclear masses, radii, and decays 
far away from stability, at present some of the required 
astrophysical inputs are still derived from often uncontrolled theoretical 
extrapolations. And even though modern experimental facilities 
of the highest intensity and longest reach will determine nuclear 
properties with unprecedented accuracy throughout the nuclear 
chart, it has been recognized that many nuclei of astrophysical 
relevance will remain beyond the experimental 
reach\,\cite{Mumpower:2015ova,RocaMaza:2008ja,Utama:2015hva}.
Thus, reliance on theoretical models that extrapolate into unknown 
regions of the nuclear chart becomes unavoidable. Unfortunately,
these extrapolations are highly uncertain and may ultimately lead
to faulty conclusions\,\cite{Blaum:2006}. However, one should not
underestimate the vital role that experiments play and will continue 
to play. Indeed, measurements of even a handful of exotic short-lived 
isotopes are of critical importance in constraining theoretical 
models and in so doing better guide the extrapolations.

Although no clear-cut remedy exists to cure such unavoidable 
extrapolations, we have recently offered a path to mitigate 
the problem\,\cite{Utama:2015hva,Utama:2016rad,Utama:2017wqe} 
primarily in the case of nuclear masses. The basic paradigm behind 
our two-pronged approach is to start with a robust underlying 
theoretical model that captures as much physics as possible 
followed by a \emph{Bayesian Neural Network} (BNN) 
refinement that aims to account for the missing 
physics\,\cite{Utama:2015hva}. Several virtues were identified
in such a combined approach. First, we observed a significant 
improvement in the predictions of those nuclear masses that 
were excluded from the training of the neural network---even 
for some of the most sophisticated mass models available in 
the literature\,\cite{Moller:1993ed,Duflo:1995,Goriely:2010bm}.
Second, mass models of similar quality that differ widely in
their predictions far away from stability tend to drastically and
systematically reduce their theoretical spread after the 
implementation of the BNN refinement. Finally, due to  the 
Bayesian nature of the approach, the refined predictions are
always accompanied by statistical uncertainties. This 
philosophy was adopted in our most recent 
work\,\cite{Utama:2017wqe}, which culminated with the 
publication of two refined mass models: the mic-mac model 
of Duflo and Zuker\,\cite{Duflo:1995} and the microscopic 
HFB-19 functional of Goriely and collaborators\,\cite{Goriely:2010bm}. 

As luck would have it, shortly after the submission of our latest 
manuscript\,\cite{Utama:2017wqe} we became aware of the newly 
published atomic mass evaluation AME2016\,\cite{AME:2016}.
This is highly relevant given that the training of the neural network 
relied exclusively on a previous mass compilation 
(AME2012)\,\cite{AME:2012}. Thus, insofar as the nearly 50 
new masses appearing in the newest compilation, ours are 
bonafide theoretical predictions. Confronting the newly refined 
mass models against the newly published data is the main goal 
of this brief report.  

This short manuscript has been organized as follows. First, no
further physics justification nor detailed account of the BNN
framework are given here, as both were extensively addressed  
in our most recent publication\,\cite{Utama:2017wqe}. Second, the 
results presented in Sec.\,\ref{results} are limited to those nuclei
appearing in the AME2016 compilation whose masses were not 
reported previously or whose values, although determined from 
experimental trends of neighboring nuclides, have a strong impact 
on $r$-process nucleosynthesis. As we articulate below, the main
outcome from this study is the validation of the novel BNN approach.
Indeed, we conclude that the improvement reported in 
Ref.\,\cite{Utama:2017wqe} extends to the newly determined nuclear 
masses---which in the present case represent true model predictions. 
We end the paper with a brief summary in Sec.\,\ref{conclusions}.

%%%%%%%%%%%%%%%%%%%%%%%%%%%%%%%%
\section{Results}
\label{results}
%%%%%%%%%%%%%%%%%%%%%%%%%%%%%%%%

\begin{figure}[h]
\vspace{-0.05in}
\includegraphics[width=0.5\columnwidth]{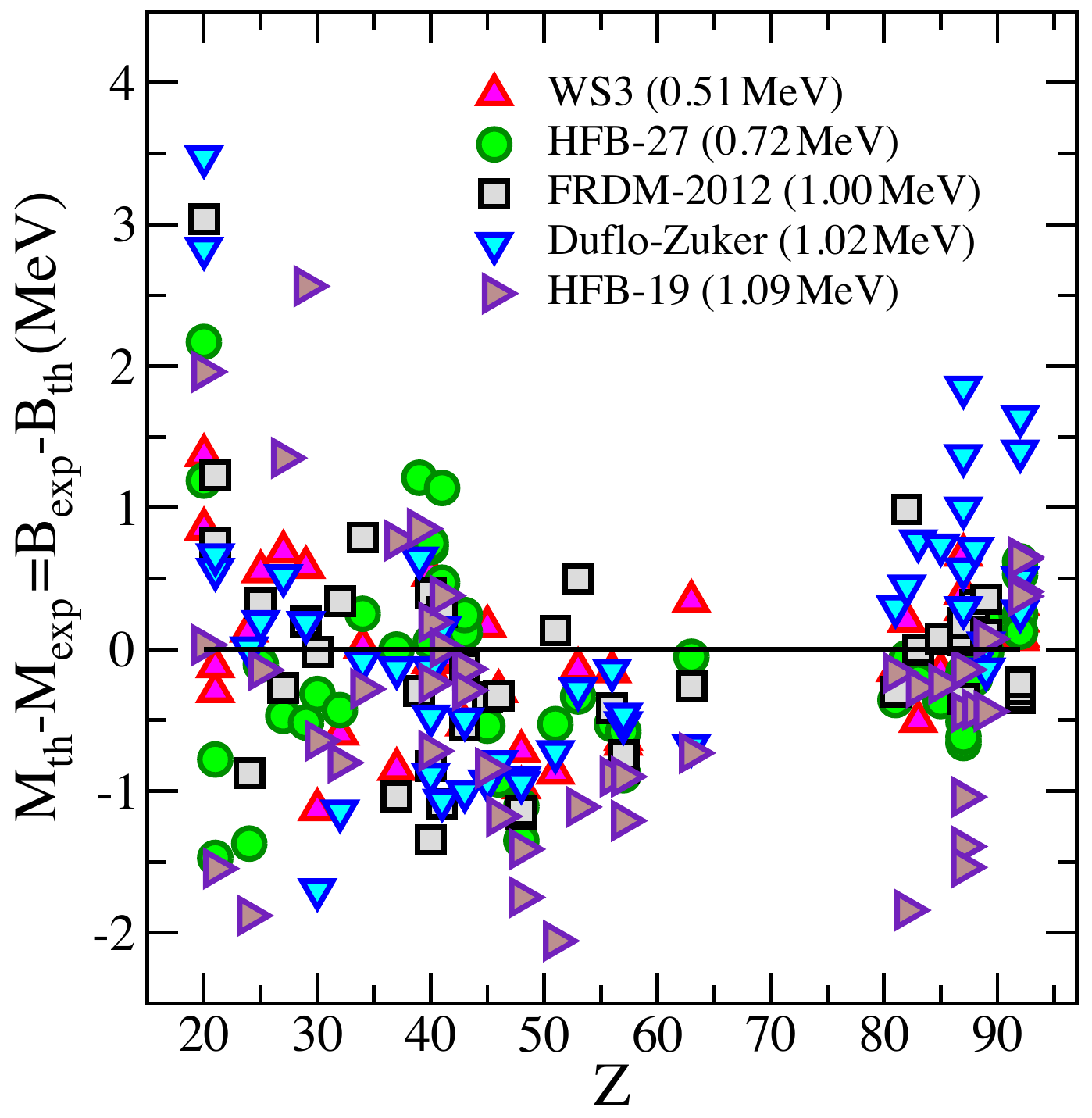}
\caption{Theoretical predictions for the total binding energy relative to
experiment for the 46 nuclei in the ${}^{40}$Ca-${}^{240}$U region that 
appear in the latest AME2016 compilation but not in any of the earlier
mass evaluations. The models shown here are representative of some 
of the most sophisticated mass models available in the literature. 
Quantities in parentheses denote the rms deviations.}
\label{Fig1}
\end{figure}

In Ref.\,\cite{Utama:2017wqe} we published refined mass 
tables with the aim of taming the unavoidable extrapolations 
into unexplored regions of the nuclear chart that are critical 
for astrophysical applications. Specifically, we refined the 
predictions of both the Duflo-Zuker\,\cite{Duflo:1995} and 
HFB-19\,\cite{Goriely:2010bm} models using the AME2012
compilation in the mass region from ${}^{40}$Ca to 
${}^{240}$U. The latest  AME2016 compilation 
includes mass values for 46 additional nuclei in the 
${}^{40}$Ca-${}^{240}$U region, and these are listed in
Table\,\ref{Table1} alongside predictions from various 
models. These include the  ``bare'' models ({\sl i.e.,} before 
BNN refinement) HFB-19\,\cite{Goriely:2010bm}, 
Duflo-Zuker\,\cite{Duflo:1995}, FRDM-2012\,\cite{Moller:2012},
HFB-27\,\cite{Goriely:2013nxa}, and WS3\,\cite{Liu:2011ama}. 
Also shown are the predictions from the BNN-improved 
Duflo-Zuker and HFB-19 models\,\cite{Utama:2017wqe}. 
The last column displays the total binding energy as reported 
in the AME2016 compilation\,\cite{AME:2016}; quantities 
displayed in parentheses in the last three columns represent 
the associated errors. Note that we quote \emph{differences} 
between the model predictions and the experimental masses 
using only the central values. Finally, the last row contains 
root-mean-square deviations associated with each of the 
models. The corresponding information in graphical form
is also displayed in Fig.\ref{Fig1}, but only for the five bare 
models discussed in the text.  

%%%%%%%%  Table I  %%%%%%%
\begin{widetext}
\begin{center}
\begin{table}[h]
\begin{tabular}{|cc||*{5}{r|}|*{2}{r|}|r|} 
 \hline\rule{0pt}{2.5ex}  
\!\!Z & N & HFB-19 & DZ\hspace{5pt} & FRDM-2012 & HFB-27 & WS3\hspace{2pt} 
       & HFB19-BNN & DZ-BNN\hspace{7pt} & AME2016\hspace{12pt}  \\               
 \hline
 \hline\rule{0pt}{2.5ex}   
 \!\!20 &  33 &  1.959 &  3.476 &  4.571 &  2.169 &  1.370 &  1.948(1.520) &  1.675(0.951) &  441.521(0.044) \\
     20 &  34 &  0.031 &  2.829 &  3.035 &  1.191 &  0.850 &  0.470(1.500) &  0.840(0.880) &  445.365(0.048) \\
     21 &  35 & -1.547 &  0.563 &  1.227 & -0.777 & -0.296 & -0.216(0.928) & -0.226(0.686) &  460.417(0.587) \\
     21 &  36 & -2.563 &  0.667 &  0.752 & -1.473 & -0.121 & -0.953(0.975) & -0.117(0.611) &  464.632(1.304) \\
     24 &  40 & -1.880 &  0.008 & -0.872 & -1.370 &  0.130 & -0.130(0.793) & -0.192(0.498) &  531.268(0.440) \\
     25 &  37 & -0.146 &  0.195 &  0.327 & -0.106 &  0.549 &  0.555(1.060) &  0.001(0.416) &  529.387(0.007) \\
     27 &  25 &  1.351 &  0.519 & -0.274 & -0.469 &  0.689 &  0.416(1.290) &  0.141(0.650) &  432.946(0.008) \\
     29 &  27 &  2.564 &  0.188 &  0.199 & -0.516 &  0.579 &  1.444(0.953) &  0.460(0.513) &  467.949(0.015) \\
     30 &  52 & -0.645 & -1.697 & -0.018 & -0.315 & -1.130 & -0.505(0.896) & -0.498(0.638) &  680.692(0.003) \\
     32 &  54 & -0.800 & -1.143 &  0.338 & -0.430 & -0.596 & -0.759(0.734) & -0.469(0.557) &  718.498(0.438) \\
     34 &  57 & -0.279 & -0.084 &  0.790 &  0.251 &  0.015 &  0.465(0.824) &  0.297(0.504) &  758.470(0.433) \\
     37 &  63 &  0.768 & -0.132 & -1.038 & -0.002 & -0.846 &  1.144(0.886) &  0.212(0.496) &  824.432(0.020) \\
     39 &  66 &  0.851 &  0.642 & -0.293 &  1.211 &  0.788 &  0.555(0.868) &  0.895(0.492) &  868.247(1.337) \\
     40 &  42 & -0.718 & -0.132 &  0.395 &  0.722 & -0.092 & -0.858(0.709) & -0.313(0.522) &  694.185(0.011) \\
     40 &  66 & -0.239 & -0.882 & -1.344 &  0.051 & -0.155 & -0.581(0.826) & -0.671(0.536) &  882.816(0.433) \\
     40 &  67 &  0.191 & -0.473 & -0.823 &  0.751 &  0.525 & -0.079(0.838) & -0.196(0.492) &  886.717(1.122) \\
     41 &  43 &  0.379 &  0.152 &  0.263 &  1.139 &  0.101 &  0.309(0.833) &  0.018(0.531) &  707.133(0.013) \\
     41 &  69 &  0.011 & -1.066 & -1.091 &  0.471 & -0.062 & -0.014(0.855) & -0.583(0.501) &  908.079(0.838) \\
     43 &  71 & -0.138 & -1.003 & -0.540 &  0.122 & -0.532 &  0.141(0.731) & -0.290(0.522) &  945.090(0.433) \\
     43 &  72 & -0.289 & -0.494 & -0.129 &  0.241 &  0.189 &  0.121(0.690) &  0.441(0.495) &  950.881(0.789) \\
     45 &  76 & -0.839 & -0.930 & -0.336 & -0.539 &  0.161 & -0.061(0.611) &  0.175(0.510) &  997.674(0.619) \\
     46 &  77 & -1.179 & -0.794 & -0.326 & -0.919 & -0.295 & -0.420(0.651) &  0.018(0.501) & 1017.214(0.789) \\
     48 &  81 & -1.411 & -0.911 & -1.165 & -1.351 & -0.976 & -0.458(0.801) & -0.560(0.444) & 1066.705(0.017) \\
     48 &  83 & -1.750 & -0.940 & -1.141 & -1.110 & -0.720 & -0.378(0.753) & -0.253(0.607) & 1075.009(0.102) \\
     51 &  87 & -2.058 & -0.724 &  0.133 & -0.528 & -0.872 & -0.395(0.824) & -0.208(0.570) & 1128.163(1.064) \\
     53 &  88 & -1.112 & -0.281 &  0.498 & -0.332 & -0.138 & -0.052(0.663) &  0.034(0.596) & 1156.518(0.016) \\
     56 &  93 & -0.899 & -0.150 & -0.415 & -0.529 & -0.157 & -0.139(0.650) & -0.074(0.420) & 1211.935(0.438) \\
     57 &  93 & -0.899 & -0.520 & -0.836 & -0.579 & -0.669 & -0.501(0.715) & -0.677(0.398) & 1222.234(0.435) \\
     57 &  94 & -1.209 & -0.460 & -0.745 & -0.889 & -0.635 & -0.690(0.713) & -0.581(0.388) & 1227.485(0.435) \\
     63 &  74 & -0.732 & -0.680 & -0.261 & -0.052 &  0.340 & -0.497(0.666) & -0.152(0.418) & 1116.629(0.004) \\
     81 & 109 & -0.167 &  0.302 & -0.298 & -0.357 & -0.149 & -0.187(0.470) & -0.225(0.289) & 1494.552(0.008) \\
     82 & 133 & -1.841 &  0.444 &  0.988 & -0.061 &  0.200 & -0.749(0.522) &  0.257(0.317) & 1666.838(0.052) \\
     83 & 111 & -0.269 &  0.764 &  0.000 & -0.219 & -0.506 & -0.280(0.457) &  0.247(0.301) & 1516.930(0.006) \\
     85 & 113 & -0.244 &  0.736 &  0.076 & -0.364 & -0.162 & -0.238(0.484) &  0.225(0.304) & 1538.336(0.006) \\
     87 & 110 & -0.443 &  0.995 &  0.061 & -0.233 &  0.282 & -0.287(0.680) &  0.394(0.475) & 1511.731(0.054) \\
     87 & 111 & -1.042 &  0.566 & -0.347 & -0.662 & -0.141 & -0.914(0.650) &  0.016(0.404) & 1520.483(0.032) \\
     87 & 115 & -0.144 &  0.293 &  0.126 & -0.164 &  0.099 & -0.113(0.571) & -0.215(0.317) & 1559.246(0.007) \\
     87 & 145 & -1.391 &  1.367 &  0.099 & -0.511 &  0.399 &  0.086(0.655) &  0.144(0.421) & 1758.409(0.014) \\
     87 & 146 & -1.538 &  1.850 &  0.183 & -0.618 &  0.667 &  0.082(0.746) &  0.426(0.586) & 1763.633(0.020) \\
     88 & 113 & -0.426 &  0.712 &  0.311 & -0.206 &  0.112 & -0.320(0.712) &  0.167(0.425) & 1541.551(0.020) \\
     89 & 116 &  0.075 &  0.042 &  0.354 &  0.215 &  0.271 &  0.153(0.739) & -0.473(0.389) & 1570.884(0.051) \\
     89 & 117 & -0.438 & -0.141 &  0.103 & -0.018 &  0.149 & -0.376(0.703) & -0.649(0.361) & 1579.583(0.050) \\
     92 & 123 &  0.408 &  0.268 & -0.346 &  0.248 &  0.319 &  0.443(0.747) & -0.215(0.437) & 1638.434(0.089) \\
     92 & 124 &  0.375 &  0.505 & -0.328 &  0.125 &  0.201 &  0.384(0.698) &  0.005(0.456) & 1648.362(0.028) \\
     92 & 129 &  0.652 &  1.641 & -0.285 &  0.622 &  0.095 &  0.504(0.677) &  0.978(0.536) & 1687.265(0.051) \\
     92 & 130 &  0.639 &  1.399 & -0.236 &  0.529 &  0.071 &  0.457(0.697) &  0.713(0.531) & 1695.584(0.052) \\
 \hline 
 \hline
    \multicolumn{2}{|c||}{$\bm{\sigma}_{\rm rms}$} &  {\bf 1.093} & {\bf 1.018} & {\bf 0.997} & {\bf 0.723} & {\bf 0.513}  
    & {\bf 0.587}\phantom{xxx}  & {\bf 0.479}\phantom{xxx}   & \\
\hline                
\end{tabular}
\caption{Theoretical predictions for the total binding energy of the 46 nuclei in the
${}^{40}$Ca-${}^{240}$U region that appear in the latest AME2016\,\cite{AME:2016}
compilation but not in AME2012\,\cite{AME:2012}. The model predictions are relative 
to the new experimental values listed in the last column and the quantities in parentheses 
represent the associated error. The last row displays the root-mean-square deviation
of each of the models. All quantities are given in MeV.}
\label{Table1}
\end{table}
\end{center}
\end{widetext}
%%%%%%%%%%%%%%%%%%%%

The trends displayed in Table\,\ref{Table1} and even more clearly illustrated 
in Fig.\,\ref{Fig1} are symptomatic of a well-known problem, namely, that 
theoretical mass models of similar quality that agree in regions where masses 
are experimentally known differ widely in regions where experimental data is 
not yet available\,\cite{Blaum:2006}. Given that sensitivity studies suggest that 
resolving the $r$-process abundance pattern requires mass-model uncertainties 
of the order of $\lesssim\!100$\,keV\,\cite{Mumpower:2015hva}, 
the situation depicted in Fig.\,\ref{Fig1} is particularly dire. However, despite the 
large scattering in the model predictions, which worsens as one extrapolates 
further into the neutron drip lines, significant progress has been achieved in 
the last few years. Indeed, in the context of density functional theory, the 
HFB-27 mass model of Goriely, Chamel, and Pearson predicts a rather small 
rms deviation of $\sim\!0.5$\,MeV for all nuclei with neutron and proton numbers 
larger than 8\,\cite{Goriely:2013nxa}. Further, in the case of the Weizs\"acker-Skyrme 
WS3 model of Liu, Wang, Deng, and Wu, the agreement with experiment is even 
more impressive: the rms deviation relative to 2149 known masses is a mere 
$\sim\!0.34$\,MeV\,\cite{Liu:2011ama}. Although not as striking, the success of 
both models extends to their \emph{predictions} of the 46 new masses listed in 
Table\,\ref{Table1}: $\sigma_{\rm rms}\!=\!0.72\,{\rm MeV}$ and 
$\sigma_{\rm rms}\!=\!0.51\,{\rm MeV}$, respectively. This represents a 
significant improvement over earlier mass models that typically predict a 
rms deviation of the order of 1\,MeV; see Table\,\ref{Table1} and
Fig.\,\ref{Fig1}.

%%%%%%%%%%%%%%%%  Figure 1 %%%%%%%%%%%%%%%
%%%%%%%%%%%%%%%%  Figure 1 %%%%%%%%%%%%%%%

\begin{figure}[h]
\vspace{-0.05in}
\includegraphics[width=0.5\columnwidth]{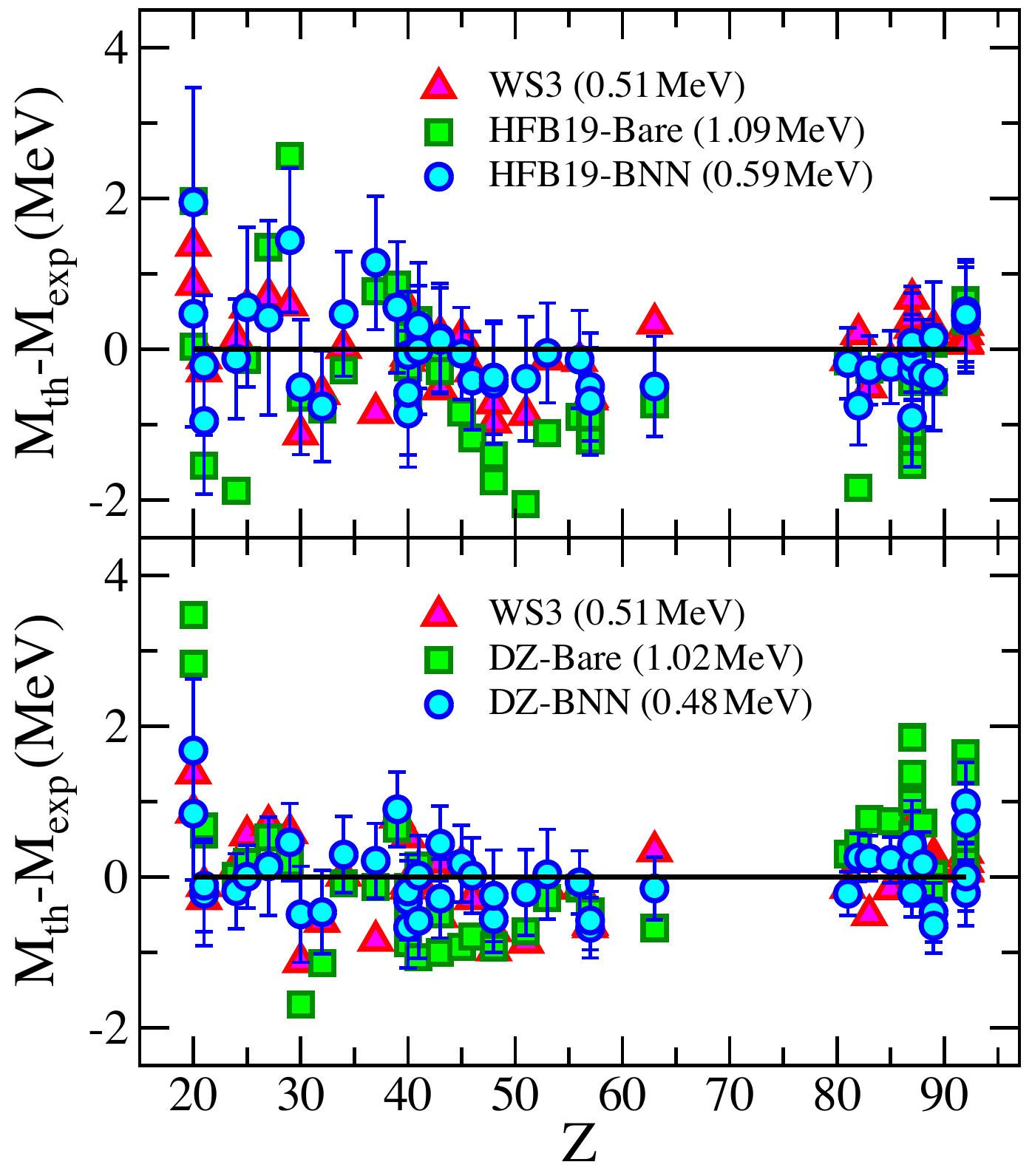}
\caption{Theoretical predictions for the total binding energy relative to
experiment for the 46 nuclei in the ${}^{40}$Ca-${}^{240}$U region that 
appear in the latest AME2016 compilation but not in any of the earlier
mass evaluations. The models shown include HFB-19 and Duflo-Zuker 
together with their corresponding BNN refinements (shown with error bars). 
For reference the WS3 model of Liu and collaborators is also shown. 
Quantities in parentheses denote the rms deviations.}
\label{Fig2}
\end{figure}

However, our main focus is to assess the improvement in the predictions 
of two of these earlier mass models (HFB-19 and DZ) as a result of the BNN 
refinement. In agreement with the nearly a factor-of-two improvement reported 
in Ref.\,\cite{Utama:2017wqe}, we observe a comparable gain in the predictions 
of the 46 nuclear masses listed in Table\,\ref{Table1}; that is,
$\sigma_{\rm rms}\!=\!(1.093\!\rightarrow\!0.587) \,{\rm MeV}$ and 
$\sigma_{\rm rms}\!=\!(1.018\!\rightarrow\!0.479) \,{\rm MeV}$ for 
HFB-19 and DZ, respectively. Of course, an added benefit of the BNN approach 
is the supply of theoretical error bars. Indeed, when such error bars are taken into 
account---as we do in Fig.\,\ref{Fig2}---then \emph{all} of the refined predictions 
are consistent with the experimental values at the 2$\sigma$ level. For reference,
also included in Fig.\,\ref{Fig2} are the impressive predictions of the WS3 model,
albeit without any estimates of the theoretical uncertainties. 

%%%%%%%%%%%%%%%%  Figure 2 %%%%%%%%%%%%%%%
%%%%%%%%%%%%%%%%  Figure 2 %%%%%%%%%%%%%%%

%%%%%%%%  Table II  %%%%%%%
\begin{widetext}
\begin{center}
\begin{table}[h]
\begin{tabular}{|cc||r|r|r|r||r|}
 \hline\rule{0pt}{2.5ex}  
\!\!Z & N & WS3\hspace{2pt} & FRDM\hspace{3pt} & DZ\hspace{5pt} & DZ-BNN\hspace{8pt} & AME2016  \\               
 \hline
 \hline\rule{0pt}{2.5ex}  
       \!48 & 84	& -1.101  & -1.704 & -1.384 &	-0.542(0.761)  & 1078.176 \\
	48 & 85	& -1.090  & -1.273 & -1.524 &	-0.556(0.954)  & 1079.827 \\
	48 & 86	& -1.252  & -1.322 & -1.664 &	-0.611(1.170)  & 1082.988 \\	
	49 & 84	& -0.910  & -0.775 & -0.676 &	-0.198(0.574)  & 1092.595 \\
	49 & 85	& -0.806  & -0.256 & -0.738 &	-0.049(0.695)  & 1094.914 \\
	49 & 86	& -1.162  & -0.590 & -1.124 &	-0.243(0.849)  & 1097.820 \\
	49 & 87	& -1.124  & -0.178 & -1.223 &	-0.190(1.030)  & 1099.832 \\	
	49 & 88	& -1.487  & -0.531 & -1.607 &	-0.476(1.240)  & 1102.439 \\
	50 & 86	& -0.785  & -0.190 & -0.445 &	0.135(0.631)   & 1114.520 \\
	50 & 88	& -1.259  & -0.246 & -1.043 &	-0.054(0.848)  & 1119.594 \\ 
 \hline 
 \hline
    \multicolumn{2}{|c||}{$\bm{\sigma}_{\rm rms}$} &  
    {\bf 1.117} & {\bf 0.877} & {\bf 1.210} & {\bf 0.369}\phantom{xxx} & \\
\hline         
\end{tabular}
\caption{Theoretical predictions for the total binding energy relative to the
experimental masses that have been estimated from experimental trends 
of neighboring nuclides\,\cite{AME:2016} and that have been identified as 
impactful in r-process nucleosynthesis\,\cite{Mumpower:2015hva}.}
\label{Table2}
\end{table}
\end{center}
\end{widetext}
%%%%%%%%  Table II  %%%%%%%

%%%%%%%%%%%%%%%%  Figure 3 %%%%%%%%%%%%%%%
\begin{figure}[h]
\vspace{-0.05in}
\includegraphics[width=0.5\columnwidth]{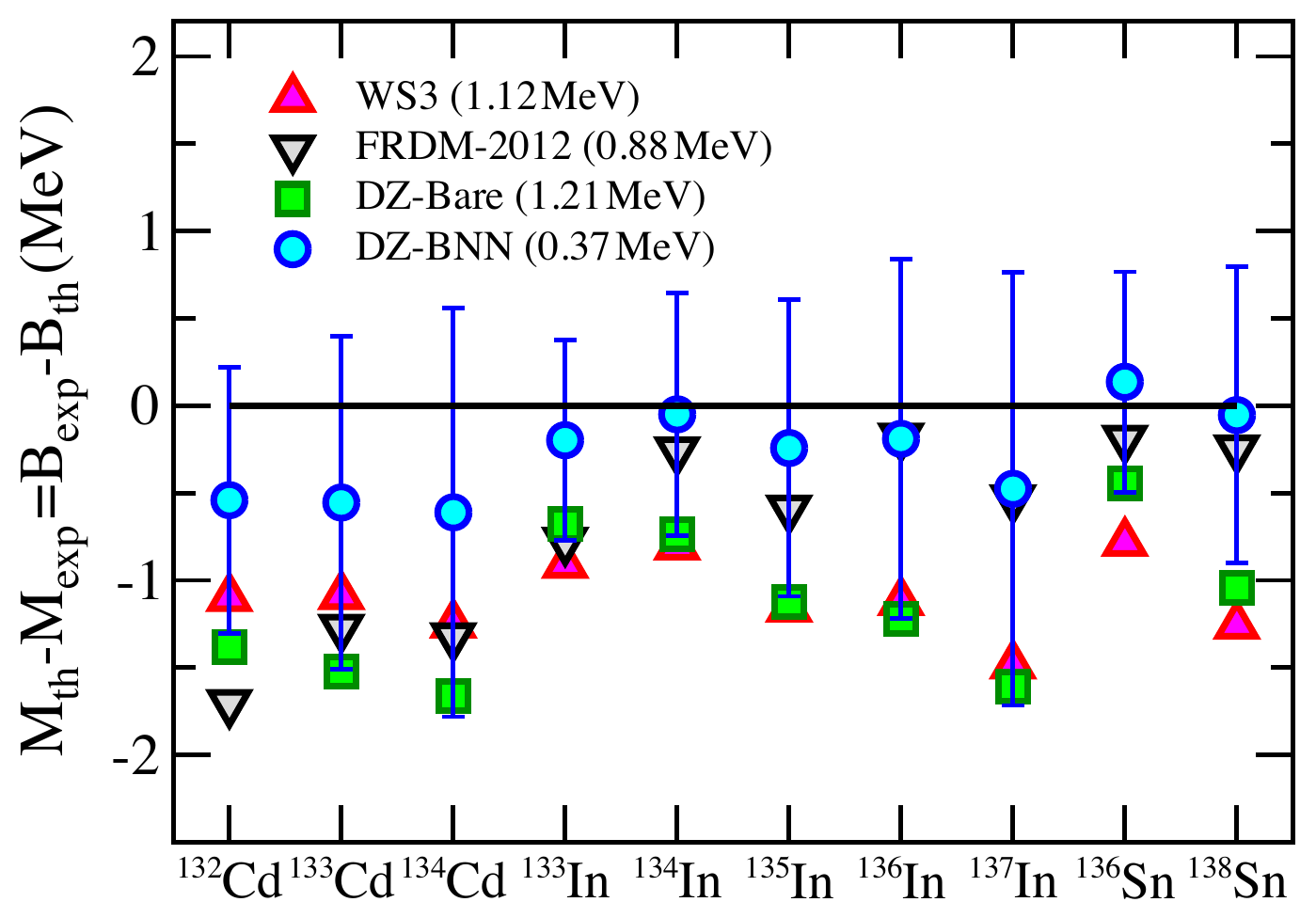}
\caption{Theoretical predictions for the total binding energy of those
nuclei that have been identified as impactful in $r$-process 
nucleosynthesis\,\cite{Mumpower:2015hva}. All experimental values 
have been estimated from experimental trends of neighboring 
nuclides\,\cite{AME:2016}. Quantities in parentheses denote the rms 
deviations.}
\label{Fig3}
\end{figure}
%%%%%%%%%%%%%%%%  Figure 3 %%%%%%%%%%%%%%%

We close this section by addressing a particular set of nuclear masses 
that have been identified as ``impactful" in sensitivity studies of the 
elemental abundances in $r$-process nucleosynthesis. These include a 
variety of neutron-rich isotopes in palladium, cadmium, indium, and tin; 
see Table I of Ref.\,\cite{Mumpower:2015hva}. As of today, none of these 
critical isotopes have been measured experimentally. However, many 
of them have been ``flagged" (with the symbol ``{\bf \#}") in the AME2016 
compilation to indicate that while not strictly determined experimentally, 
the provided mass estimates were obtained from \emph{experimental 
trends of neighboring nuclides}\,\cite{AME:2016}. In Table\,\ref{Table2} 
theoretical predictions are displayed for those isotopes that have been
both labeled as impactful and flagged. Predictions are provided for the
WS3\,\cite{Liu:2011ama}, FRDM-2012\,\cite{Moller:2012}, 
DZ\,\cite{Duflo:1995}, and BNN-DZ\,\cite{Utama:2017wqe} mass models. 
Root-mean-square deviations of the order of 1\,MeV are recorded for all
models, except for the improved Duflo-Zuker model where the deviation
is only 369\,keV. This same information is depicted in graphical form in 
Fig.\,\ref{Fig3}. The figure nicely encapsulates the spirit of our two-prong 
approach, namely, one that starts with a mass model of the highest quality 
(DZ) that is then refined via a BNN approach. The improvement in the 
description of the experimental data together with a proper assessment
 of the theoretical uncertainties are two of the greatest virtues of the BNN 
approach. Indeed, the BNN-DZ predictions are consistent with all masses 
of those impactful nuclei that have been determined from the experimental 
trends.

%%%%%%%%%%%%%%%%%%%%%%%%%%%%%%%%
\section{Conclusions}
\label{conclusions}
%%%%%%%%%%%%%%%%%%%%%%%%%%%%%%%%
Nuclear masses of neutron-rich nuclei are paramount to a variety 
of astrophysical phenomena ranging from the crustal composition
of neutron stars to the complexity of $r$-process nucleosynthesis.
Yet, despite enormous advances in experimental methods and 
tools, many of the masses of relevance to astrophysics lie well 
beyond the present experimental reach, leaving no option but to 
rely on theoretical extrapolations that often display large systematic
variations. The current situation is particularly troublesome 
given that sensitivity studies require mass-model uncertainties 
to be reduced to about $\lesssim\!100$\,keV in order to resolve 
$r$-process abundances. 

There are at least two different approaches currently used to alleviate 
this problem. The first one consists of painstakingly difficult measurements 
near the present experimental limits that aim to inform and constrain 
mass models. The second approach is based on a global refinement 
of existing mass models through the training of an artificial neural network.
This is the approach that we have advocated in this short contribution. Given 
that the training of the neural network relied exclusively on the AME2012 
compilation, our approach was validated by comparing our theoretical
predictions against the new information provided in the most recent 
AME2016 compilation. 

The comparison against the newly available AME2016 data was highly 
successful. For the nearly 50 new mass measurement reported in the 
${}^{40}$Ca-${}^{240}$U region, the rms deviation of the two BNN-improved 
models explored in this work (Duflo-Zuker and HFB-19) was reduced by 
nearly a factor of two relative to the unrefined bare models. Further, for the 
masses of several impactful isotopes for the $r$-process, the predictions 
from the improved Duflo-Zuker model were fully consistent with the new
AME2016 estimates and in far better agreement than some of the most 
sophisticated mass models available in the literature. Finally and as
important, all nuclear-mass predictions in the BNN approach incorporate
properly estimated statistical uncertainties. When these theoretical error
bars are incorporated, then \emph{all} of our predictions are consistent 
with experiment at the 2$\sigma$ level.

Ultimately, improvements in mass models require a strong synergy between
theory an experiment. Next-generation rare-isotope facilities will produce new 
exotic nuclei that will help constrain the physics of weakly-bound nuclei. In
turn, improved theoretical models will suggest new measurements on a few
critical nuclei that will best inform nuclear models. We are confident that 
the BNN approach advocated here will play a critical role in this endeavor,
particularly in identifying those nuclei that have the strongest impact in 
resolving some outstanding questions in nuclear astrophysics. We
are hopeful that in the near future mass-model uncertainties---both statistical 
and systematic---will be reduced to less than $100$\,keV, which represents 
the elusive standard required to resolve the $r$-process abundance pattern. 

%%%%%%%%%%%%%%%%%%%%%%%%%%%%%%%%%%%%%%%%%%%%

\begin{acknowledgments}
 We are thankful to Pablo Giuliani for many stimulating discussions. 
 This material is based upon work supported by the U.S. Department 
 of Energy Office of Science, Office of Nuclear Physics under Award 
 Number DE-FG02-92ER40750.
\end{acknowledgments}
\vfill\eject

%%%%%%%%%%%%%%%%%%%%%%%%%%%%%%%%%%%%%%%%%%%%

%\bibliography{../ReferencesJP.bib}
\bibliography{AME2016vsBNN.bbl}
\end{document}